\PassOptionsToPackage{expansion=false}{microtype}

\documentclass[manuscript,nonacm]{acmart}

\AtBeginDocument{%
  }

\setcopyright{none}
\renewcommand\footnotetextcopyrightpermission[1]{}

\usepackage{etoolbox}
\usepackage{xspace}
\usepackage[most]{tcolorbox}
\usepackage{tabularx}

\def\tsc#1{\csdef{#1}{\textsc{\lowercase{#1}}\xspace}}
\tsc{WGM}
\tsc{QE}
\tsc{EP}
\tsc{PMS}
\tsc{BEC}
\tsc{DE}

\begin{document}

\title{Lilo: A Multi-Agent Conversational Sandbox for Children's Digital Communication}

\author{Ofri Elbaum}
\affiliation{%
  \institution{University of Haifa}
  \city{Haifa}
  \country{Israel}
}

\author{Amit Malik}
\affiliation{%
  \institution{University of Haifa}
  \city{Haifa}
  \country{Israel}
}
\email{amalik05@campus.haifa.ac.il}

\author{Tomer Malkevich}
\affiliation{%
  \institution{University of Haifa}
  \city{Haifa}
  \country{Israel}
}

\author{Zohar Elyoseph}
\affiliation{%
  \institution{University of Haifa}
  \city{Haifa}
  \country{Israel}
}

\author{Adir Solomon}
\affiliation{%
  \institution{University of Haifa}
  \city{Haifa}
  \country{Israel}
}
\email{asolomon@is.haifa.ac.il}

\author{Rotem Dror}
\affiliation{%
  \institution{University of Haifa}
  \city{Haifa}
  \country{Israel}
}

\author{Oshrat Ayalon}
\affiliation{%
  \institution{University of Haifa}
  \city{Haifa}
  \country{Israel}
}
\email{oayalon@is.haifa.ac.il}
\authorsaddresses{}

\renewcommand{\shorttitle}{Lilo: A Multi-Agent Conversational Sandbox}
\renewcommand{\shortauthors}{Elbaum et al.}

\begin{abstract}

Children's peer conflicts increasingly play out in group chats, where tone and timing are stripped away, and a hurtful message is easier to send than to say aloud. We present Lilo, a multi-agent conversational system that lets children practice such conversations, so they are better prepared when they encounter them in real digital communication. In Lilo, the child joins two AI-driven peers mid-argument, coached by a mentor character, also named Lilo, through a Stop-Think-Act traffic light. Through an exploratory study with 28 children (aged 9-11), we examine how children engage with Lilo's simulated peer conflict and the mentor's traffic-light interventions. The study included an intervention with Lilo, surveys before and after the intervention, and exit interviews. Participants reported significantly higher self-efficacy for coping with unpleasant messages after their interaction with Lilo. They felt safe experimenting, knowing nothing they wrote would be held against them.

\end{abstract}


\maketitle

\section{Introduction}\label{sec:intro}


Children enter the digital sphere at a young age. In the US, for example, two-thirds of children receive their first smartphone by age 10 or younger \cite{martin2025life}. A similar picture emerges for social media (SM) use, where a substantial share of 8-12 year-olds already use SM platforms despite the official minimum age of 13 \cite{rideout2022commonsensecensus}. Accordingly, peer-to-peer communication also takes place through these channels, and so do the conflicts that arise within it. This is a demanding arena for children in middle childhood, who are still developing the social and emotional skills needed to navigate it \cite{seo2024chacha}.

In particular, text-based messaging is a lean medium, as it strips away tone of voice, facial expression, and timing cues, such as a pause or an instant reply. These and other cues normally help people tell how a message is meant \cite{walther1992interpersonal, daft1986organizational, rhoads2010face}. Online disinhibition makes a hurtful message easier to send than to say aloud, and it removes the visible reaction that would normally restrain it \cite{braddock2023introduction, margolis2023social}. As a result, the skills that keep a disagreement from escalating, such as pausing before replying, considering how a message will land, reframing, and answering assertively rather than aggressively, must be exercised exactly where practice is most costly. A misjudged message in a real group chat cannot be taken back.

The field of social skills training has long grappled with this challenge \cite{ding2024technology}. Prior work has used digital approaches to build such skills, for example, an automated social skills trainer that provided real-time feedback on non-verbal cues such as speaking volume and eye contact \cite{tanaka2016teaching}. However, prior work largely relied on scripted or rule-based interaction rather than dynamic, responsive dialogue \cite{fu2022social}. Moreover, most of it focused on clinical populations, particularly children on the autism spectrum \cite{ding2024technology, ali2020virtual, hurst2020social}. In recent years, the rise of generative AI has begun to narrow this gap, with LLM-based role-play helping improve conflict skills and self-efficacy \cite{shaikh2024rehearsal, lin2024imbue}. Yet this work has targeted adults, and child-facing LLM agents so far focus on emotion disclosure or collaborative learning, not rehearsal of a peer conflict of one's own \cite{seo2024chacha, liu2024peergpt}.

\begin{figure*}[]
 \centering
  \includegraphics[width=\textwidth]{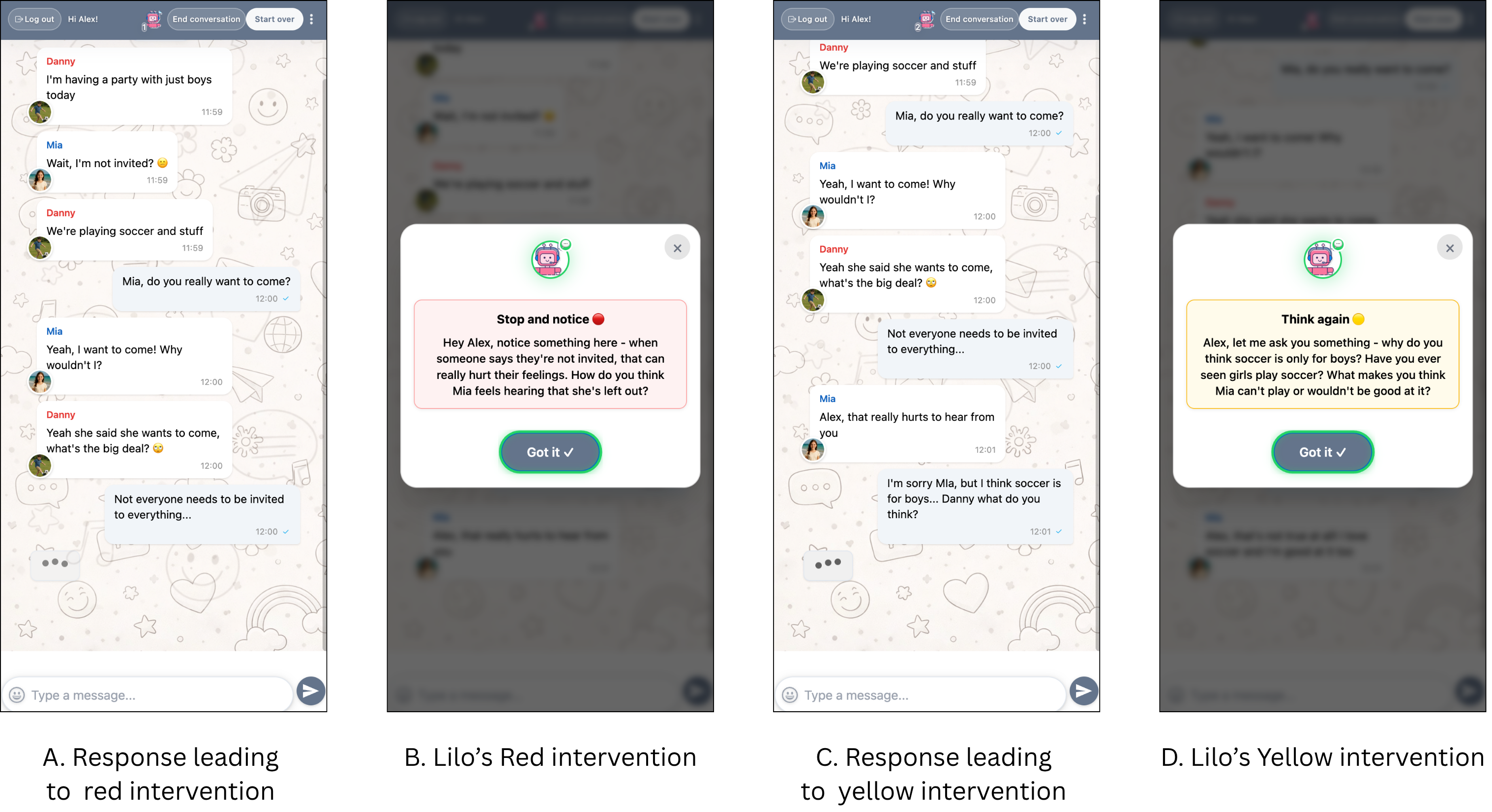}
  \caption{Examples of Lilo's traffic-light interventions in the chat interface. \textbf{A} A child (Alex) response that leads to a Red intervention. \textbf{B} Lilo's Red intervention, prompting the child to notice the impact of what was said. \textbf{C} A child response that leads to a Yellow intervention. \textbf{D} Lilo's Yellow intervention, asking the child to consider the other side's perspective. Screens show an English-localized build; the study itself was conducted in Hebrew. Danny and Mia correspond to Sagi and Reut.}
  \label{fig:lilo_first_screens}
\end{figure*}

We present \emph{Lilo} (Figure~\ref{fig:lilo_first_screens}\footnote{For readability, figures throughout the paper use illustrative English-language screenshots with the placeholder names Danny and Mia in place of Sagi and Reut; the system itself was designed and evaluated in Hebrew, in which Sagi and Reut are common children's names.}), a multi-agent conversational system in which children practice digital communication inside a simulated peer conflict. 
From the user's point of view, the system is designed around three AI-based characters: 1) Sagi, a boy who pushes back and resists compromise; 2) Reut, a girl who registers the emotional cost of what is said; and 3) Lilo, the system's namesake, who acts as a mentor rather than a peer. The conflict itself is pre-defined. Sagi, despite being a good friend of Reut, refuses to invite her to his birthday party, insisting it will be a ``boys only'' party. The user of the system is a child, who joins the ongoing chat and is asked to participate the way they would in a regular group chat with their friends. The system's interface resembles WhatsApp, a platform widely used by children. Behind the scenes, an orchestrator agent reads each turn and decides which characters respond, how firmly they hold their positions, and whether the mentor should intervene. The interventions operationalize the Stop-Think-Act model \cite{kendall1977efficacious, durlak2011impact} through a traffic-light metaphor \cite{elias1997promoting}, in which Lilo draws the child's attention to a problematic issue, suggests other perspectives to consider, and provides positive feedback as the child communicates appropriately.

In this study, we explored the potential of an LLM-based conversational system to help children practice digital communication. We evaluated Lilo with 28 children aged 9-11 who regularly use WhatsApp with their classmates. Each session paired the interaction with Lilo with surveys administered before and after it, a semi-structured exit interview, and the system's interaction logs. We asked how children's self-efficacy and intended coping strategies changed following the session, how they engaged with the system and its interventions, and how they experienced it.

Participants reported significantly greater perceived ability to cope with unpleasant behavior after their session with Lilo, both when directed at them and when another child was the target, while their perceived ease of identifying such behavior did not significantly change. The exit interview discussions point to two possible reasons behind this gain, echoing two of Bandura's sources of self-efficacy \cite{bandura1977self}. Participants described \textit{actively} resolving the conflict themselves, and recalled \textit{Lilo telling} them explicitly when they had done well. The characters' credibility, in turn, depended on several factors: their register, persona, and conduct in the chat. For example, whether a character behaved as expected (conduct) influenced whether the child perceived it as more AI-like or more human. We discuss what these findings mean for designing simulated peers and mentoring agents for children in Section~\ref{sec:discussion}.

\section{Related Work}\label{sec:background}
\subsection{ Children’s Digital Communication} 
\label{sec:children-digital-comm}


Digital platforms offer children real benefits, including maintaining friendships beyond school, supporting self-presentation, and enabling the self-expression and emotional communication that strengthen their sense of belonging \citep{d2021social, uhls2017benefits}. They also expose children to risks, chief among them here cyberbullying, as well as the Fear of Missing Out (FOMO), which pressures children to stay constantly connected to avoid social exclusion \citep{braddock2023introduction, uhls2017benefits, d2021social}.

Focusing on peer-to-peer communication, digital environments often serve as a volatile social arena for children in middle childhood, given their limited social experience and still-developing communication skills \cite{seo2024chacha}, so minor misunderstandings can rapidly intensify into significant social conflict \cite{d2021social}. This volatility is theoretically grounded in the Social Information Processing perspective \cite{walther1992interpersonal} and Media Richness Theory \cite{daft1986organizational}, which posit that text-based communication is inherently ``lean'': lacking the tone, facial expression, and body language of face-to-face interaction, it leaves children struggling to accurately interpret a peer's intent \cite{rhoads2010face}.

Other effects further exacerbate this digital social environment.  Online disinhibition leads children to be more outspoken and less restrained online, often sharing aggressive or hurtful remarks they would not communicate in person. In part, this happens because the medium hides the victim's immediate reaction, filtering out the social cues that would normally trigger empathy and restrain aggression \cite{braddock2023introduction, margolis2023social}. Although children may use paralinguistic adaptations like emojis to ``filter in'' social information, these lean media often remain inefficient for resolving peer friction due to high equivocality, making it particularly difficult for children with maturing emotional regulation to reach a shared understanding \cite{brody2019cues}.

\subsection{Employing technology for social skills training}
\label{sec:tech-sst}
Technology has long been employed to support social and communicative development across a wide range of populations, from adults with social anxiety to individuals with clinical diagnoses \cite{ding2024technology}. Early automated systems, such as the ``automated social skills trainer'' \cite{tanaka2016teaching}, demonstrated that structured, repeatable practice through human-agent interaction could meaningfully improve social communication skills, providing real-time feedback on non-verbal cues such as speaking volume, pitch, and eye contact. Within this broader landscape, a substantial portion of technology-supported social skills training focused on children, particularly those with Autism Spectrum Disorder (ASD) \cite{ding2024technology}. Systems designed for this population ranged from conversational agents to physically embodied robots, each offering a controlled and safe environment for social practice. The LISSA system \cite{ali2020virtual}, for instance, engaged teenagers with ASD in multi-topic conversations while providing real-time feedback on non-verbal behavior. Embodied platforms extended this further, where the robot Moxie \cite{hurst2020social} leveraged physical presence to sustain engagement and deliver therapeutic content, and Keepon, a robot mediator \cite{shen2018stop}, demonstrated that a robotic agent could intervene in live peer conflicts and prompt more constructive resolution among young children.

Despite their promise, early social skills training (SST) systems shared significant interaction and architectural limitations. A systematic review of commercially available voice-based conversational agents for children's social-emotional learning (SEL) found that existing systems overwhelmingly provided shallow experiences, relying on a small set of rigid interaction styles such as ``The Lecturer'' - systems that talked at users without interaction - or ``The Bulldozer'' - systems that ignored user input and continued along a fixed script \cite{fu2022social}. At the architectural level, Al Owayyed et al.'s ARTES framework argues that most existing systems emphasize either social situation simulation or educational scaffolding at the expense of the other, limiting their overall training impact \cite{al2025agent}. Early efforts to address these gaps introduced more structured pedagogical sequences, such as embedding social stories within phased conversation flows or leveraging story-based interactions to support socioemotional strategy acquisition \cite{gagan2022designing, fu2023self}. Yet these systems remained largely rule-based and scripted, constraining the naturalness and adaptability of interaction. The emergence of large language models (LLMs) offered a new trajectory, one in which conversational agents could move beyond fixed scripts toward dynamic, context-sensitive dialogue while still operating within structured pedagogical frameworks.

A parallel line of work, meanwhile, has shifted focus from clinical populations toward neurotypical children navigating online peer conflicts such as cyberbullying \cite{piccolo2021chatbots, morag2022tobe, zou2024social}. Early efforts approximated peer interaction through structured virtual role-play: FearNot!, for instance, immersed children in animated bullying episodes in which they took on the role of an ``invisible friend'' advising a virtual victim, demonstrating that simulated social scenarios could foster empathy and bystander intervention \cite{vannini2011fearnot}. More recent participatory design studies have surfaced clearer expectations for next-generation systems: children and educators envision chatbots that are trustworthy, non-judgmental, and act as ``understanding friends'' - humanized but not pretending to be human - offering empathetic, contextualized ``learning at the moment'' guidance during online conflicts, while complementing rather than replacing parents and trusted adults \cite{piccolo2021chatbots, zou2024social}. Notably, these studies stop short of deploying actual LLM-driven systems, instead articulating socio-technical requirements that anticipate this shift.

The emergence of LLMs has transformed role-play from a scripted exercise into a generative practice tool, capable of producing dynamic, context-sensitive dialogue. Recent systems such as Rehearsal \cite{shaikh2024rehearsal} and IMBUE \cite{lin2024imbue}, both designed for adults, demonstrate that grounding LLM outputs in established social-science frameworks can support meaningful skill acquisition. Both systems combine theory-grounded prompting with just-in-time feedback, and both improved outcomes in randomized studies - Rehearsal participants used more cooperative strategies in an actual conflict than a control group given didactic materials, and IMBUE raised participants' self-efficacy and skill use relative to simulation without feedback. Within the children's domain, recent work has likewise begun exploring LLM-based agents for communication and learning \cite{seo2024chacha, liu2024peergpt, malik2026multi}. ChaCha \cite{seo2024chacha}, for example, applies a similar logic by pairing an LLM with a state machine that structures the dialogue into pedagogical phases, guiding children through emotion-coaching conversations. Together, this wave of work suggests that LLMs can be effective educational tools when their generative flexibility is shaped by an established pedagogical or psychological framework. However, as Hiniker et al. \cite{hiniker2021can} caution, children are quick to learn routines from conversational agents but simultaneously value the ``ontological distinction'' between humans and machines, adapting whether and how they transfer such routines based on context. This suggests that generative agents should be designed with attention to children’s existing interpersonal norms and to how they distinguish interactions with humans from those with machines.

\subsection{Psychological and Pedagogical Background}
\label{sec:psych-pedagogy}
The pedagogical intervention in this study was based on the Stop, Think, Act model, an evidence-based strategy for fostering self-regulation and responsible decision-making, grounded in the principles of cognitive-behavioral therapy (CBT) and social-emotional learning (SEL) \cite{kendall1977efficacious,durlak2011impact}. The model was implemented using an intuitive ``traffic light'' user interface, a visual metaphor commonly used to scaffold self-regulation and impulse control in children \cite{elias1997promoting} (p.~29). The ``stop'' phase (red light) was designed to promote impulse control by encouraging an intentional pause before responding.
The ``thinking'' phase (yellow light) formed the core of the intervention, guiding participants through a structured cognitive process that included: (a) social perspective taking, i.e., the ability to infer the thoughts and feelings of others involved in the situation \cite{gehlbach2023understand}, a particularly critical skill in digital contexts lacking nonverbal cues \cite{walther1992interpersonal,gehlbach2023understand}; and (b) cognitive reappraisal, in which users practice reframing the situation to regulate their emotional responses, a strategy positively associated with interpersonal functioning and well-being \cite{gross2003individual}. 
Finally, the ``action'' phase (green light) focused on implementing the chosen solution through assertive communication, as opposed to passive or aggressive responses, which is a key goal of social skills training \cite{sert2003assertiveness}. This phase's design also incorporated praise for effective communication in the moment. The overall aim of the process is to equip users with the practical skills required for responsible digital citizenship \cite{prasetiyo2023promoting}.

\section{Lilo: System Design and Framework}\label{sec:lilogeneral}
Lilo is designed as a multi-agent conversational system that embeds structured pedagogical scaffolding within open-ended, peer-like dialogue, and adapts to users' communication skill development in real time. Building on the design principles established in Section~\ref{sec:background}, and in particular, the value of grounding generative agents in psychological and pedagogical frameworks \citep{seo2024chacha, shaikh2024rehearsal, lin2024imbue}, this section describes our system in detail.

\subsection{Conversational Design and Educational Foundations}\label{sec:conversationaldesign}


Lilo addresses the gap identified in Sections~\ref{sec:children-digital-comm}–\ref{sec:tech-sst} by giving neurotypical children aged 9–11 a safe, simulated peer environment in which to practice impulse control, social perspective-taking, cognitive reappraisal, and assertive communication (Section~\ref{sec:psych-pedagogy}) \citep{kendall1977efficacious, gehlbach2023understand, gross2003individual, sert2003assertiveness}. Within this framing, we made three design choices: (1) exposing children to a realistic peer conflict; (2) embedding the Stop-Think-Act framework within the conversation; and (3) adapting the interaction to the child's evolving communication. LLM-driven simulation suits these directions because, unlike scripted agents, LLMs sustain context-sensitive, generative dialogue \citep{shaikh2024rehearsal, lin2024imbue, seo2024chacha} and, when shaped by an established pedagogical framework, do so within the structure required for measurable skill acquisition.

\paragraph{Scripted Scenario Framework.}

Lilo's interactions unfold within a controlled scenario: a peer conflict over a ``boys only'' birthday party invitation, designed to elicit empathy, perspective-taking, and conflict resolution. Anchoring the interaction in a fixed scenario follows a methodological tradition in technology-mediated social skills training, in which structured conversational contexts allow rehearsal without real-world interpersonal consequences  \citep{vannini2011fearnot, shaikh2024rehearsal, lin2024imbue, haroon2024twips}. 
This design choice serves three functions. First, a fixed scenario gives every child the same conflict and supports repeated practice of the same target skills \citep{durlak2011impact}.  Second, the target communication skills, such as proposing compromises and taking the other side's perspective, are also specified in the Orchestrator's prompt and inform its per-turn assessment of the child's recent communication and difficulty adjustments. Third, holding the situation constant across participants means that differences in how children responded are not confounded with differences in what they were responding to.

\paragraph{Psychological Foundations and Persona Design.}

Tseng et al.  \citep{tseng2024two} distinguish two paradigms for persona in LLMs: role-playing, in which the persona is assigned to the LLM, and personalization, in which the LLM models the user. Lilo's design combines both. On the role-playing side, three character agents - Sagi, Reut, and Lilo - are assigned stable identities, communication styles, and behavioral tendencies, and interact within a multi-agent setting (Section~\ref{sec:architecture}). On the personalization side, the orchestrator continuously models the individual child's communication conduct and skill development, conditioning character behavior - tone, willingness to compromise, level of challenge, and timing of interventions - on this evolving user model. While Tseng et al. \cite{tseng2024two} treat role-playing and personalization as distinct lines of research, they note that the two can co-occur in a single system. Lilo is one such case, and the integration of the two is central to the system's pedagogical function.

Lilo's mentor design draws on ChaCha's \citep{seo2024chacha} use of an LLM-driven agent for emotion coaching with children, whose reflective, non-instructive approach, Seo et al. argue, sustains disclosure and self-reflection at this age. Lilo adopts the same orientation, and applies it within a simulated peer conflict. The mentor prompts the child to consider interpretations and feelings (reflective) rather than supplying a correct answer (non-instructive), while the peer agents continue the conflict around that guidance.

\paragraph{Traffic Light Intervention Model.}
Lilo operationalizes the Stop-Think-Act framework introduced in Section~\ref{sec:psych-pedagogy} as a system-level intervention protocol: the mentor agent (Lilo) moves through Red, Yellow, and Green phases of guidance, each scaffolding the corresponding stage of the child's reasoning. 
The phase is not pre-scheduled but is selected by the orchestrator based on the conversational context, the user's recent communication patterns, and the emotional state of the exchange.  The orchestrator may also select no mentor intervention, allowing the peer exchange to continue when additional guidance is unnecessary. This contextual timing is intended to preserve conversational flow rather than interrupt every turn, avoiding the rigid interaction styles described in earlier child-facing systems \citep{fu2022social}.

\subsection{Architectural Framework}\label{sec:architecture}
Lilo's architecture is built on multi-agent orchestration: specialized persona agents are coordinated by a central orchestrator that manages conversation flow and maintains contextual coherence across the interaction. This design is motivated by recent work on Chain-of-Agents reasoning \citep{zhang2024chain} and on the degradation of LLM reasoning performance with longer input contexts \citep{levy2024same}. Rather than relying on a single agent driven by one long prompt that must encode every character, every rule, and every educational objective, Lilo distributes the interaction across short, focused prompts, each handled by a separate agent. This decomposition is intended to reduce the load on a single prompt and to support consistent adherence to each character's persona behavior, though it may also increase response latency because generating a turn can involve multiple model calls.

\begin{figure*}[]
  \centering
  \includegraphics[width=\textwidth]{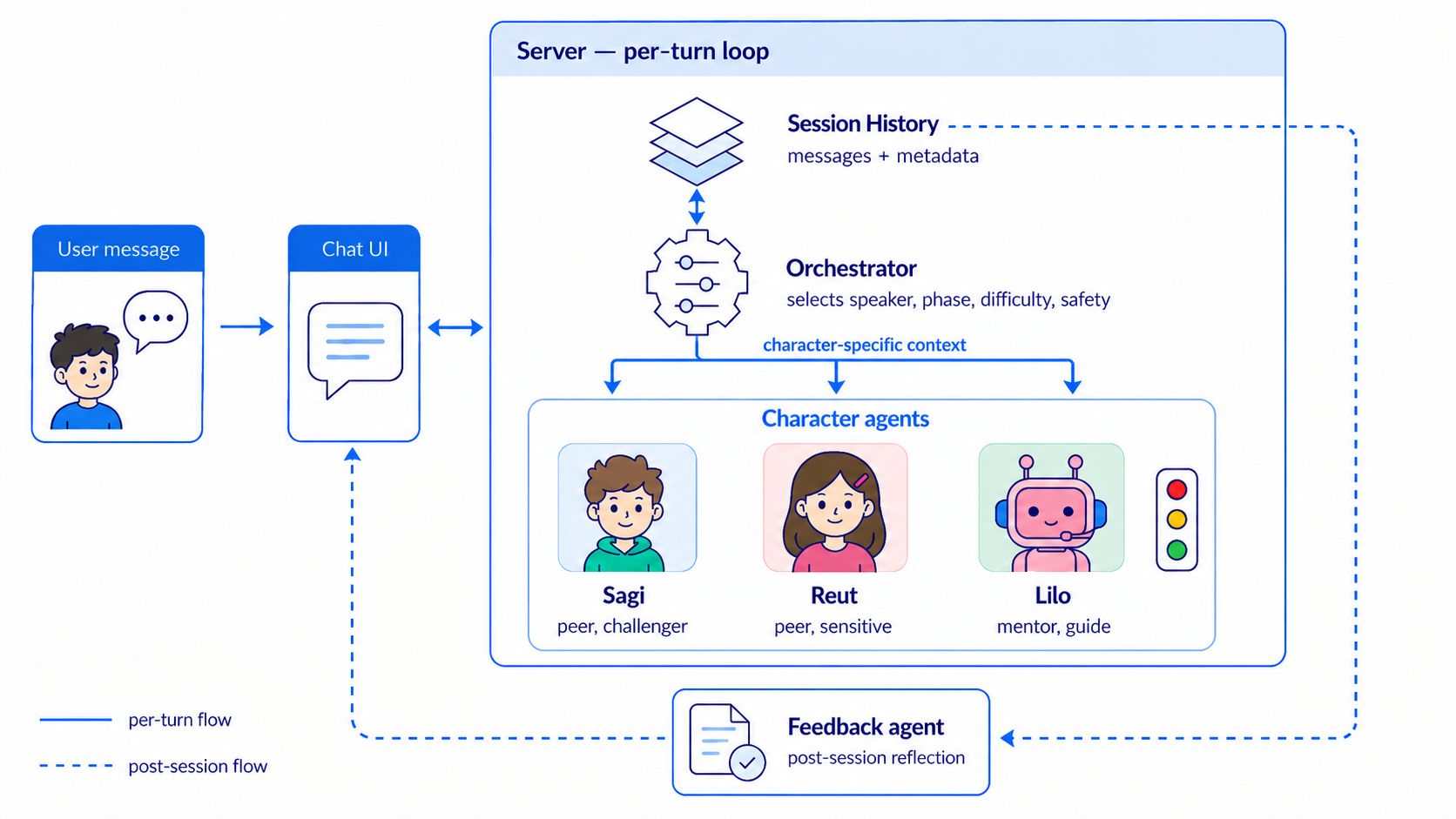}
  \caption{System architecture of the Lilo multi-agent framework. }
  \label{fig:lilo-arch}
\end{figure*}

The architecture comprises five LLM-based agents: three character agents (Sagi, Reut, and Lilo),  a central Orchestrator Agent that coordinates their interactions based on ongoing analysis of the conversation, and a Feedback Agent that generates structured feedback at the end of the session. In addition, the system is supported by session history and flow-management logic (Figure~\ref{fig:lilo-arch}, which we describe later in    Section~\ref{sec:systemcomponents}). The remainder of this subsection describes how the orchestrator manages conversation flow and adaptive difficulty, and the safety layers that operate alongside the educational logic.

\paragraph{Adaptive Orchestration.} 

The Orchestrator Agent is the system's coordinating layer: it reads the unfolding conversation and decides how the system should respond at each turn. Its design draws on ChaCha's \citep{seo2024chacha} use of structured state management for child-computer interactions, but replaces transitions over a predefined state graph with per-turn decisions produced by an LLM-based orchestrator. At each user turn, the orchestrator analyzes the recent conversation history, the child’s latest message and communication patterns, and the interaction dynamics among the agents. It selects one or more responding agents, including Sagi, Reut, and potentially the mentor agent (Lilo), determines a traffic-light phase when Lilo is selected, and assigns a difficulty level that is passed to the responding agents.

Difficulty progresses through four levels: \textit{Intense, Hard, Medium, and Low}, each characterized by more or less challenging character behavior and willingness to compromise. Initial interactions begin at Intense. After 2–3 consecutive messages at the current level that demonstrate effective communication, such as proposing compromises, considering both sides, using respectful language, or de-escalating the conflict, the orchestrator decreases the difficulty by one level. Ineffective or aggressive communication leaves the level unchanged. This adaptive calibration is intended to keep children appropriately challenged while preventing frustration that could impede learning \citep{vygotsky1978mind}.

The orchestrator thus operates not as a rule executor but as a contextual decision-maker, able to keep the conversation aligned with both immediate dialogue dynamics and longer-term pedagogical objectives. The system logs each interaction and orchestration decision in detail. The use of these logs for analysis is described in Section~\ref{sec:methodology}.

\begin{figure}[]
  \centering
  \includegraphics[width=0.4\textwidth]{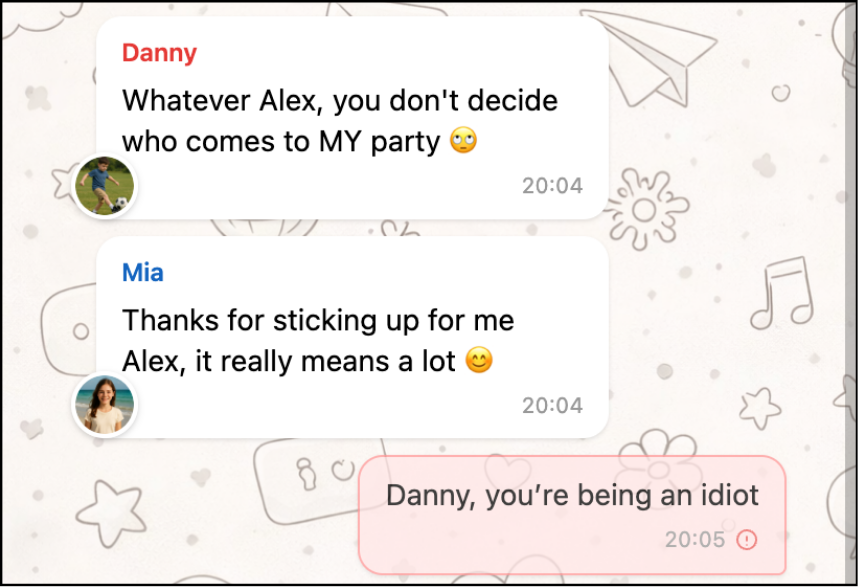}
  \caption{Example of problematic language and Lilo's response.}
  \label{fig:lilo-red-note}
\end{figure}

\paragraph{Safety Layer.}
The system's architecture includes safety as a continuous concern alongside the educational logic. Every user message is analyzed in real time for inappropriate content, with automatic flagging and intervention when problematic language or behavioral patterns such as curses, insults, humiliation, and other harmful messages are detected. This safety judgment is distinct from ordinary disagreement or impolite phrasing that remains within the simulated conflict. This detection is performed by the orchestrator agent as part of its per-turn analysis, which flags problematic language and routes the response to the mentor agent (Lilo) with a red traffic-light phase. Flagged messages are displayed to the user with a red background, signaling to the child that the message was not actually sent (Figure~\ref{fig:lilo-red-note}); the character agents (Sagi, Reut) do not see the flagged message at all, so the conversation continues without them reacting to it. For example, here are two cases exemplifying the difference:

\begin{quote}
    \textit{Safety flag, message is not sent: ``You're so annoying, everyone thinks so''}
\end{quote}

\begin{quote}
    \textit{Regular red light: ``Not everyone needs to be invited to everything''}
\end{quote}

\paragraph{Language Model and Localization.} All LLM-based agents in Lilo used Claude 3.7 Sonnet, selected during system development for its instruction-following and Hebrew-language capabilities. Agent-specific roles and behaviors were defined through role-specific prompting and orchestration, without model fine-tuning. 

Lilo was designed for Hebrew-language interaction using role-specific prompts tailored to the local linguistic and conversational context of children's digital communication. Peer-character prompts encourage short, age-appropriate messages, informal expressions, slang, moderate emoji use, and occasional ambiguity, while the mentor prompt uses child-friendly Hebrew and guides children to consider colloquial and ambiguous expressions in context. Together, these prompt-level adaptations were intended to approximate everyday peer group-chat communication while preserving each agent's distinct educational role.

\subsection{Supporting Components and Character Agents}\label{sec:systemcomponents}
\paragraph{Memory and Context Management.} 

The system maintains a session-scoped conversation history, storing message history together with metadata such as role, timestamps, content flags, and difficulty-level markers that track transitions in the system's adaptive challenge level. Before each agent call, the orchestrator constructs the relevant conversation context from this history. For example, messages flagged as inappropriate are excluded from the character agents' regular context while remaining available in the complete session history for safety review and post-session feedback. Character agents (Sagi, Reut) also receive a filtered view that excludes the mentor's interventions, preserving the illusion of independent peer actors unaware of the pedagogical scaffolding.


Retaining this history within the session also allows Lilo's interventions to reference the child's earlier behavior. The full history remains available beyond the per-turn loop: at the end of the interaction, it is passed to the Feedback Agent, which generates the end-of-conversation feedback shown to the child, situating this agent outside the main per-turn loop shown in Figure~\ref{fig:lilo-arch}.

\paragraph{Flow and Trigger Orchestration.}
The orchestrator coordinates the timing and sequencing of system responses. It manages both reactive responses to user input and proactive auto-trigger mechanisms that maintain conversation engagement during periods of user inactivity. When users remain silent for an interval of 60 seconds, the auto-trigger system initiates character-to-character exchanges, in which the characters continue talking to each other rather than waiting for the user. The mechanism is designed primarily to draw the user back into active participation. Having the characters talk to each other (rather than prompt the user directly) also has the side effect of modeling that peer conversation continues on its own, mimicking how conversations unfold among real peers even when one participant is momentarily inactive.

When multiple characters are selected to respond within the same turn, the orchestrator generates and delivers their messages sequentially, with each generation conditioned on the messages that have already been produced in that turn. This sequential coordination prevents characters from contradicting or echoing each other within a single turn. The result is a multi-character exchange that approximates realistic group chat dynamics, in which the user and the two character peers (Sagi and Reut) interact much as they might in an actual group conversation.

\paragraph{Persona Implementation.}

Sagi, Reut, and Lilo were introduced in Section~\ref{sec:conversationaldesign} as the three character agents that constitute Lilo's role-playing layer. Here, we describe the behavioral role each plays in the session:

\begin{enumerate}
    \item Sagi (The Challenger): opinionated, stubborn, and boundary-testing, holding firm on the ``boys only'' invitation despite pushback and using deliberately ambiguous or slangy messages (e.g., ``leave it'', ``you're weak''). His behavior requires users to practice patience, interpretation of ambiguous text, and de-escalation.
    \item Reut (The Sensitive Peer): emotionally expressive and direct about her feelings, with responses ranging from hurt and frustration (when excluded or dismissed) to warmth and relief (when validated). If the user remains passive, she presses them to state their position. Her behavior encourages users to consider the emotional impact of what they write, and how empathic responses repair the relationship.
    \item Lilo (The Mentor-Guide): a mentor rather than a peer, intervening through brief ($\leq$60-word) messages structured around the traffic-light phases (Section~\ref{sec:architecture}) and addressed to the child rather than to Sagi or Reut. Lilo uses hedging language (e.g., ``it might be that...'') to model non-judgmental interpretation, encourages perspective-taking and reframing, and suggests possible responses without solving the conflict. During the Green phase, Lilo reinforces the constructive strategies listed in Section~\ref{sec:architecture}.
\end{enumerate}

\subsection{Multi-Layered Prompting Architecture}\label{sec:multilayered}

\paragraph{Static Prompt Foundation.}
The static prompt layer defines the persona of each character. These prompts were developed through team discussions that included an expert educational psychologist to ensure psychological validity and educational effectiveness. Each character prompt encodes six components of character definition: 1) identity markers (character names and, for the child characters, age and gender); 2) functional role within the conversation system; 3) core personality traits and behavioral tendencies; 4) characteristic communication styles and linguistic patterns; 5) primary educational objectives; and 6) adaptive behavioral rules that govern response variation.
The static prompts were informed by emotion coaching and peer mediation research \citep{seo2024chacha, shen2018stop}, with the specific prompt content developed by the team. The prompts include contextual background and inter-character relationships, which support consistent character behavior across the session, alongside few-shot examples of appropriate responses in different conversational situations.

\paragraph{Dynamic Prompting and Contextual Adaptation.}
The dynamic prompting layer enables real-time adaptation of character responses based on the unfolding conversation and the child's evolving conduct. At each turn, the system injects current session variables into the static prompt templates: user profile information (name, gender, and age), the overall difficulty level, reasoning produced by the Orchestrator (described in the next paragraph), and character-specific conversation context (described in Section~\ref{sec:systemcomponents}); when Lilo is selected, the current traffic-light intervention phase is also supplied. This dynamic integration allows each character to remain consistent with its own persona while adapting its responses to the immediate situation: difficulty calibrates the peer characters' behavioral intensity, while the traffic-light state governs the mentor's form of intervention when it is used.

\paragraph{Orchestrator Reasoning Integration.}
The dynamic prompt layer also incorporates the Orchestrator's per-turn reasoning. When the Orchestrator selects which character or characters should respond, it produces reasoning that explains why those characters were chosen and what educational purpose their responses should serve. This reasoning is injected into the dynamic prompt alongside the session variables described above, so that character responses are shaped not only by their personas but also by the Orchestrator's pedagogical intent at that turn.


\section{Methodology}\label{sec:methodology}
\begin{figure*}[]
  \centering
  \includegraphics[width=\textwidth]{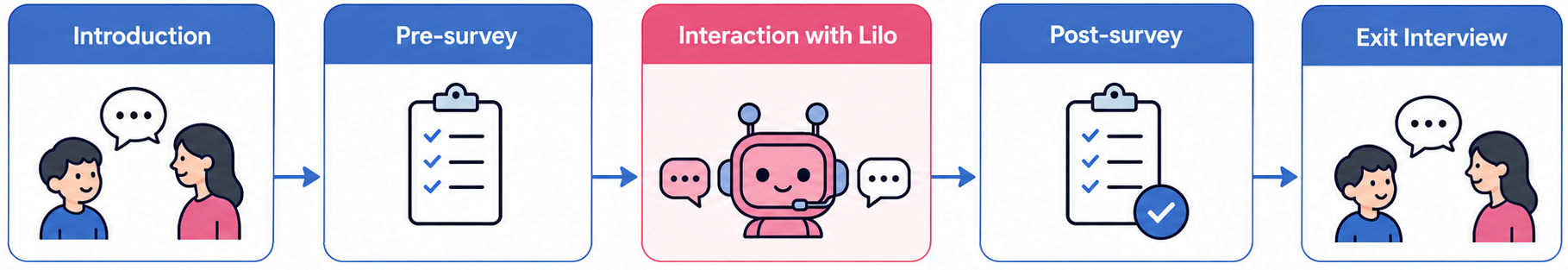}
  \caption{Study  design stages, as described in \S\ref{sec:procedure}.}
  \label{fig:study-procedure}
\end{figure*}

We conducted a user study with three objectives: to examine whether and how children's self-efficacy and intended coping strategies regarding unpleasant WhatsApp interactions changed following a structured interaction with Lilo; to characterize children's engagement with the system and the interventions it produced; and to learn about their experience with it. To address these objectives, each session combined surveys administered before and after the interaction with a brief semi-structured interview that followed. In addition, the system logged each interaction, which we analyzed descriptively. The study design, including the interview protocol and survey questions, was approved by our institution's Institutional Review Board (approval number: 297-25), and all parents provided informed consent prior to their child's participation.

\subsection{Procedure}
\label{sec:procedure}
The study was conducted entirely over Zoom, with one participant per session. Each session lasted approximately one hour and followed five stages: (i) introduction, (ii) pre-survey, (iii) interaction with Lilo, (iv) post-survey, and (v) exit interview. Figure~\ref{fig:study-procedure} outlines the procedure. See Appendix \ref{sec:instruments} for all the study-related details and instruments used. 

Recruitment proceeded in two stages, both conducted online via Qualtrics. In the first stage, parents completed a screening survey that collected relevant child information, including age, gender, and whether the child uses WhatsApp to communicate with classmates (rather than only with family). The screening also asked parents to rate, on a 1-to-5 scale, their child's reading and typing abilities in WhatsApp, since the study required reading and typing at a conversational pace. To be eligible, children had to be aged 9-11, use WhatsApp to communicate with classmates, and have parent-rated reading and typing abilities of at least 3 on the 1-to-5 scale. Parents whose children met these criteria were invited to a second stage, in which they received a detailed description of the study - including the role-play scenario, the AI-based nature of the characters, and safety mechanisms in Lilo - and provided informed consent. They also completed a brief demographic questionnaire. 
Parents were instructed not to over-explain the study to their child in advance, in order to avoid biasing the child's behavior during the session.

 Sessions were led by three of the authors, one experimenter per session. At the start of the session, the experimenter introduced themselves, briefly described what the session would include, and reminded the child that participation was voluntary and that they could stop at any time. A parent was asked to be present at the beginning of the session to help with the technical setup, such as connecting the child's phone to Zoom and sharing the phone screen so that the experimenter could see the child's view of Lilo during the interaction. After the setup, the parent was asked to leave the room so that the child could engage with the system as naturally as possible, while remaining nearby in case of technical issues.

The pre-survey was administered in two parts. Part~A was completed by the child alone, via a Qualtrics link sent during the session. It covered topics the child might prefer to answer privately rather than out loud to the experimenter: the child's prior exposure to unpleasant behavior in classroom WhatsApp groups, adapted from \cite{aizenkot2018cyberbullying}, and how they would typically respond to such behavior. Part~B was completed jointly with the experimenter, who read the questions aloud and recorded the child's answers in Qualtrics. Part~B included self-efficacy items \cite{bandura1977self},  adapted from \cite{elyoseph2024using}, such as the extent to which the child would know how to deal with unpleasant behavior directed at them on WhatsApp. Throughout the survey, the closed-ended questions used a 1-7 Likert scale. 
Each closed-ended item was followed by an open-ended follow-up asking the child for an example. 

After the pre-survey, the experimenter introduced the expected interaction with Lilo. The child was told that they were about to use an application that resembles WhatsApp, and that they would join an ongoing conversation between two AI-based peer characters, Sagi and Reut, who were currently in a disagreement. The child was asked to imagine that Sagi and Reut were their friends and to respond as they would in a real WhatsApp conversation. The experimenter explained that Lilo, an additional character, would intervene from time to time, and asked the child to read Lilo's interventions and the traffic-light feedback rather than clicking past them immediately. The experimenter remained on Zoom throughout the interaction and could assist if needed. The interaction was capped at approximately 20 minutes, after which the experimenter would end the session if it had not ended naturally.  
The post-survey was then administered jointly by the experimenter, again via Qualtrics. It repeated the items from the pre-survey, and added Lilo-specific evaluation items, such as perceived future helpfulness of the experience and willingness to use Lilo again. 

Finally, the exit interview was audio-recorded, lasting approximately 15~minutes. The interview followed a semi-structured protocol adapted from \cite{seo2024chacha} and covered the child's experience with Lilo, including their perceptions of the characters and the traffic-light. 
At the end of the interview, the experimenter clarified that Sagi was deliberately designed to behave in a confrontational manner and that the research team does not endorse such behavior. Prior to data collection, two pilot sessions were conducted to refine the study protocol, including the survey questions, procedure, and interview protocol; data from these sessions are excluded from all analyses. As compensation for participation, participants were offered the equivalent of approximately 10~USD.

\subsection{Participants}

\label{sec:participants}
We recruited the participants between July and December 2025 through various channels, including social media platforms such as Facebook and X, WhatsApp groups, and a poster located at a children's science museum. Ultimately, we recruited 28 participants. Children's ages ranged from 9 to 11 (M=9.9), and the sample included 17 girls (61\%) and 11 boys (39\%). Table~\ref{tab:demographics} presents the demographic breakdown by age and gender. All children owned a smartphone and used WhatsApp to communicate with classmates. Parents of participating children were predominantly 
highly educated (93\% held at least a bachelor's degree), and secular or traditional (72\%).

\begin{table}[h]
\centering
\caption{Participant demographics by age and gender.}
\label{tab:demographics}
\begin{tabular}{lccc}
\hline
\textbf{Age} & \textbf{Girls (\textit{n})} & \textbf{Boys (\textit{n})} & \textbf{Total (\textit{n})} \\
\hline
9  & 5 & 4 & 9  \\
10 & 10 & 6 & 16 \\
11 & 2 & 1 & 3  \\
\hline
\textbf{Total} & \textbf{17} & \textbf{11} & \textbf{28} \\
\hline
\end{tabular}
\end{table}

\subsection{Data Analysis}
\label{sec:analysis}
We report on the analysis of three data sources: 1) system logs; 2) surveys administered before (pre) and after (post) the intervention; and 3) exit interviews. We describe the analysis approach for each data source.

\paragraph{System logs.}
System log data were analyzed descriptively, where we report on characteristics of both participants' interactions and system behavior, including engagement patterns and intervention distributions. Two participants were excluded from this analysis due to technical issues that prevented complete interaction logs from being recorded, leaving n=26 for log analysis. The survey analysis below is also based on 26 participants, but not the same 26: two different participants were excluded there, for the reasons noted in that paragraph, so the two datasets overlap on 24 participants.

\paragraph{Surveys.}
Descriptive statistics were used to summarize participants' prior exposure to unpleasant behavior in class WhatsApp groups, their coping strategy selections, and their post-interaction evaluations of Lilo. We only analyzed closed-ended questions, as open-ended questions mostly covered topics that were later discussed during the exit interviews. To examine pre-post changes, we assessed three items individually: ease of identifying unpleasant WhatsApp behavior, self-efficacy for coping when personally targeted, and self-efficacy for intervening as a bystander. Each item captures a distinct construct and was therefore analyzed separately rather than combined into a composite score. Two participants were excluded from the survey analysis: one did not complete the post-survey, and one left a full item block in the pre-survey unanswered. This leaves 26 participants for all survey analyses. We first assessed normality of difference scores using the Shapiro-Wilk test. Difference scores for two of the three items showed non-normal distributions ($p < .05$). Therefore, Wilcoxon signed-rank tests were used consistently across all three items for comparability. 

\paragraph{Exit interviews.}
Qualitative data from the exit interviews were analyzed using thematic analysis \citep{terry2017thematic}. The interviews were conducted in Hebrew, and quotations included in the paper were translated into English and attributed using anonymized participant identifiers. The interviews were analyzed by the same three authors who conducted the sessions (Section~\ref{sec:procedure}), together with a fourth senior researcher. The three researchers began by independently coding an initial set of two interviews. One of these interviews was part of the pilot stage (that is, excluded from the analysis). The team then met to discuss their codes and develop an initial codebook, which was further refined across two additional iterations of coding and discussion. Once the codebook was finalized, two of the three researchers independently analyzed the remaining interviews, 10 and 6 respectively, using the agreed-upon codebook, and were instructed to stop coding once they no longer identified new themes in the data. The codebook is presented in Appendix~\ref{sec:codebook}.
Coding stopped after 17 of the 28 recorded interviews, at which point the two researchers were no longer identifying new themes. Prior work also reported a similar phenomenon, where most codes were established after analyzing a relatively small number of participants \cite{guest2006many}. Guest et al. further identified two conditions under which saturation is reached early: participants who are homogeneous with respect to the research domain, and a structured guide in which the same questions are asked of everyone. These two conditions hold in our case, where every child encountered the same fixed scenario with the same characters, every interview followed the same semi-structured protocol, and participants fell within a narrow age range (9-11). 
To assess inter-rater reliability, two interviews were independently coded by both researchers. Agreement was assessed at the level of the five overarching themes, treating each paragraph-theme pair as a unit. The coders agreed on 68.6\% of units, with Cohen's $\kappa = .63$, indicating substantial agreement
\citep{landis1977measurement}.

\section{Results}\label{sec:results}


\subsection{Participants' Prior Experiences}

When asked about their prior experience with instant messaging, most participants reported that their class had WhatsApp groups with classmates. 17 participants (65.4\%) reported that there was a WhatsApp group that included all students in their class, while 22 (84.6\%) reported that there were WhatsApp groups that included only some students from their class. These categories were not mutually exclusive, as participants could report the existence of both types of groups.

\begin{figure*}[]
  \centering
  \includegraphics[width=\textwidth]{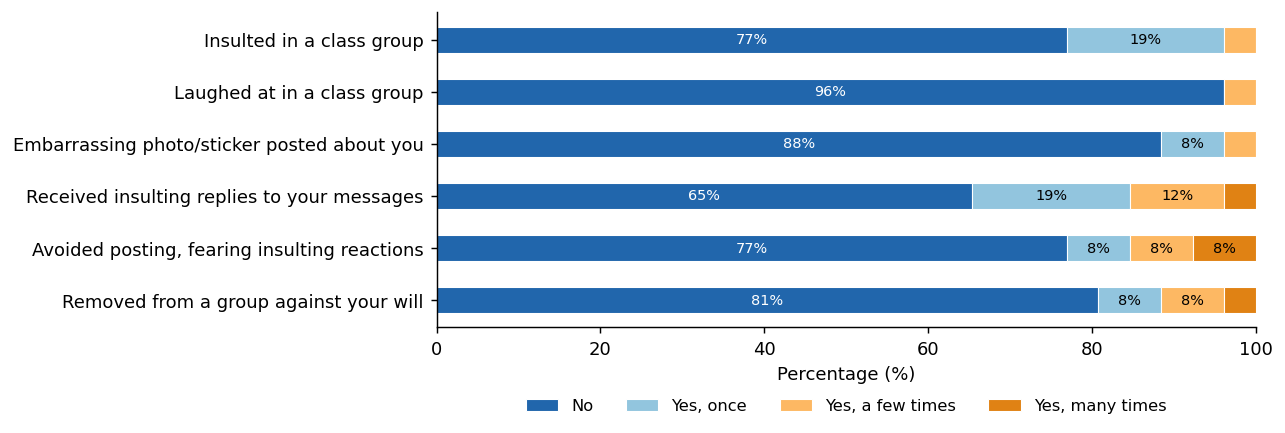}
  \caption{Prior experiences with unpleasant behavior in class WhatsApp groups in the past 6 months}
  \label{fig:whatsapp_experience}
\end{figure*}

Overall, the majority of participants did not report any of the surveyed negative experiences during the preceding six months, as presented in Figure~\ref{fig:whatsapp_experience}. Exposure varied across experience types, and participants could report more than one type of experience. The most frequently reported experience was receiving insulting replies to messages, reported by 9 participants (34.6\%). Six participants (23.1\%) reported having been insulted in a class group, and six reported that they had avoided participating because they feared insulting responses. Five participants (19.2\%) reported having been removed from a group against their will, three (11.5\%) reported that an embarrassing photograph or sticker had been posted about them, and one (3.8\%) reported having been laughed at in a class group.

\subsection{System Logs}
\label{sec:logs}

We report on 26 sessions that were conducted during the data collection, one per participant. Altogether, participants sent 261 messages, with a mean of 10.0 messages per session (median 9). 
As for session duration, sessions lasted 17.8 minutes on average (median 16.8, and longest 31.4).  Across all sessions, 936 discussion messages were exchanged in total. Children's 261 messages accounted for 28\% of these, while the three agents produced the remaining 675: Lilo 181 (19\%), Sagi 270 (29\%), and Reut 224 (24\%). 

\begin{figure*}[]
  \centering
  \includegraphics[width=\textwidth]{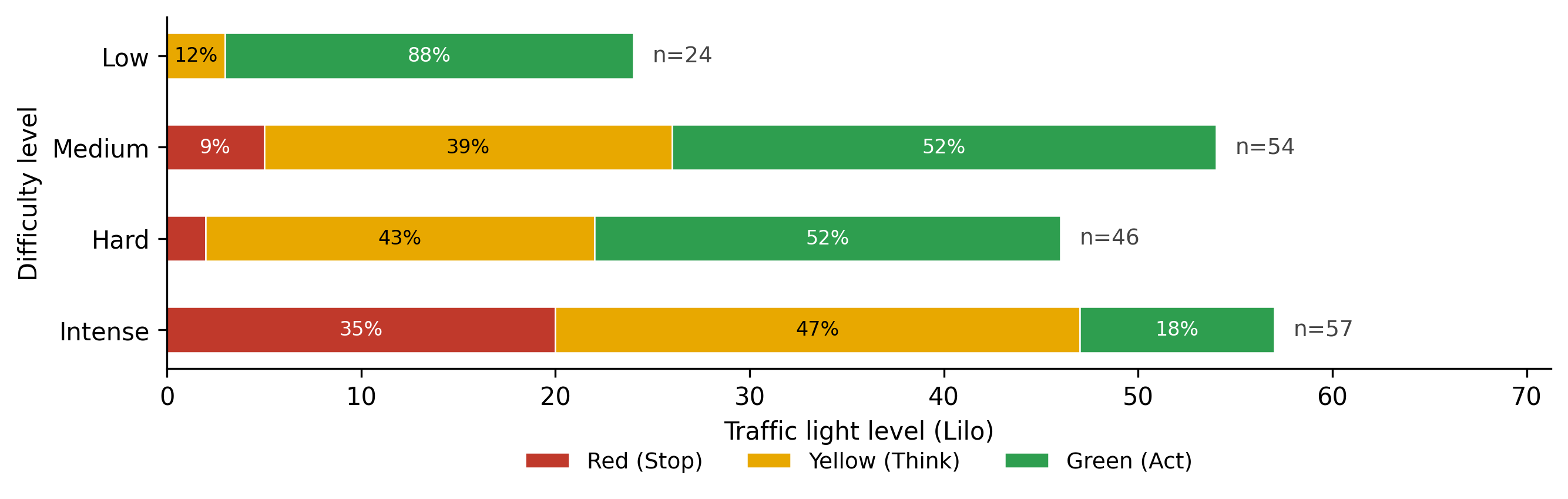}
  \caption{Traffic light level by difficulty level}
  \label{fig:traffic_difficult}
\end{figure*}

As described earlier (\S\ref{sec:architecture}), the orchestrator manages two aspects of each turn: Lilo's traffic-light intervention (Green/Yellow/Red) and the session's difficulty level (Low/Medium/Hard/Intense). Across the 26 sessions, the orchestrator produced 181 traffic-light decisions - Green (Act) was most frequent (83, 46\%), followed by Yellow (Think, 71, 39\%) and Red (Stop, 27, 15\%) - and assigned a difficulty level to 265 turns, spread across all four levels (47 Low, 81 Medium, 52 Hard, 85 Intense). Four turns were auto-generated to foster discussion, exceeding the 261 participants' messages. Taking the two aspects together, Figure~\ref{fig:traffic_difficult} cross-tabulates traffic-light phase against difficulty level; of the 265 turns carrying a difficulty label, 181 also carried a traffic-light decision, and the cross-tabulation is restricted to these. The proportion of Green decisions decreased as difficulty increased, from 88\% at Low to 52\% at both Medium and Hard, and 18\% at Intense. Red decisions were rare overall (27 of 181, 15\%) and heavily concentrated at the Intense level: 20 of the 27 Red decisions (74\%) occurred there. Within Intense-level turns specifically, Red accounted for 35\% of decisions - compared with 9\% at Medium, 4\% at Hard, and none at Low. 

\subsection{Changes in Coping Strategies and Perceived Ability}
\label{subsec:changes_coping}

The post-intervention survey included two parts: questions about participants' anticipated coping with negative WhatsApp experiences, and questions specific to their experience with Lilo (\S\ref{subsec:postsurvey}). In this subsection, we report on the first part of the survey. We asked which strategies participants would use in response to unpleasant behavior directed at them or witnessed toward another child (Figure~\ref{fig:coping_strategies}, Appendix~\ref{sec:coping}). We observed that selections were largely unchanged before and after the intervention. For example, referring to the scenario in which the child is personally insulted,  the same strategies - resolving the situation privately and seeking outside help - remained the most common for both pre- and post- intervention.

\begin{figure*}[]
  \centering
  \includegraphics[width=\textwidth]{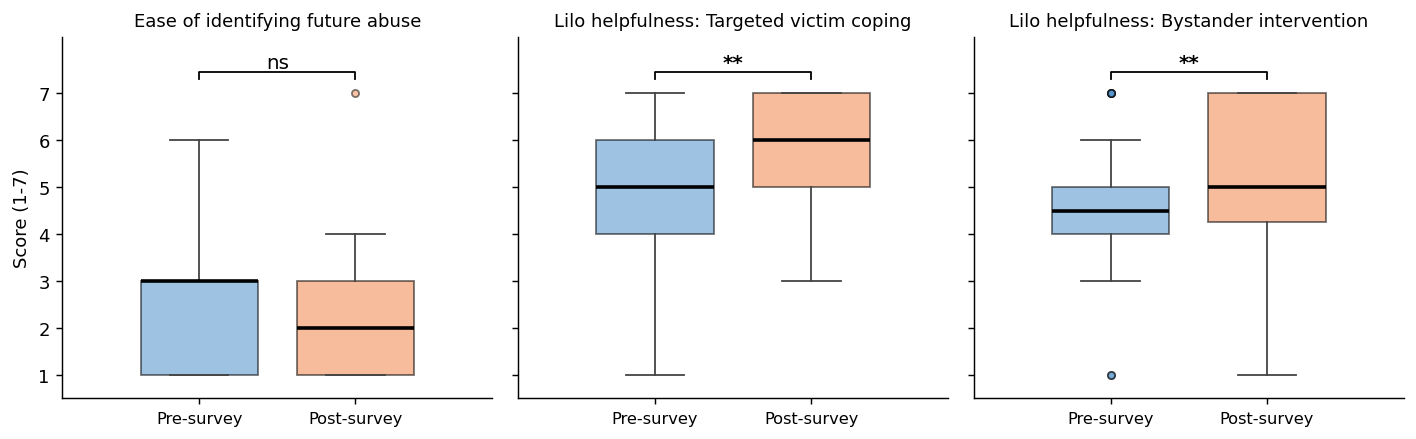}
  \caption{Pre–Post comparison of perceived self-efficacy and identification of unpleasant interaction.  }
  \label{fig:self_efficacy}
  \vspace{2pt}
  {\footnotesize Note: $^{*}p<.05$, $^{**}p<.01$, Wilcoxon signed-rank test.}
\end{figure*}

We compared responses to three individual survey questions before and after the interaction with Lilo using Wilcoxon signed-rank tests (Figure~\ref{fig:self_efficacy}). The questions assessed participants' perceived ease of identifying unpleasant behavior on WhatsApp, their perceived ability to cope when such behavior was directed toward them, and their perceived ability to respond when another child was treated unpleasantly. For the identification question, lower scores indicated greater perceived ease. For the two coping questions, higher scores indicated greater perceived ability.

As for ease of identifying future unpleasant behavior, participants' responses showed a small descriptive improvement, but the pre-post difference was not statistically significant. Scores decreased from M=2.50 (SD=1.39) before the interaction to M=2.23 (SD=1.50) afterward (Wilcoxon test: W=49.0, p=.314).
In contrast, participants reported significantly greater perceived ability to cope when unpleasant behavior was directed toward them. Scores increased from M=4.85 (SD=1.57) before the interaction to M=5.77 (SD=1.18) afterward (Wilcoxon test: W=11.0, p=.001.)
Participants also reported greater perceived ability to respond when another child was treated unpleasantly. Scores increased from M=4.58 (SD=1.53) before the interaction to M=5.38 (SD=1.47) afterward (Wilcoxon test: W=9.0, p=.006.)

\subsection{Post-Interaction Evaluation of Lilo} 
\label{subsec:postsurvey}

\begin{figure*}[]
  \centering
  \includegraphics[width=\textwidth]{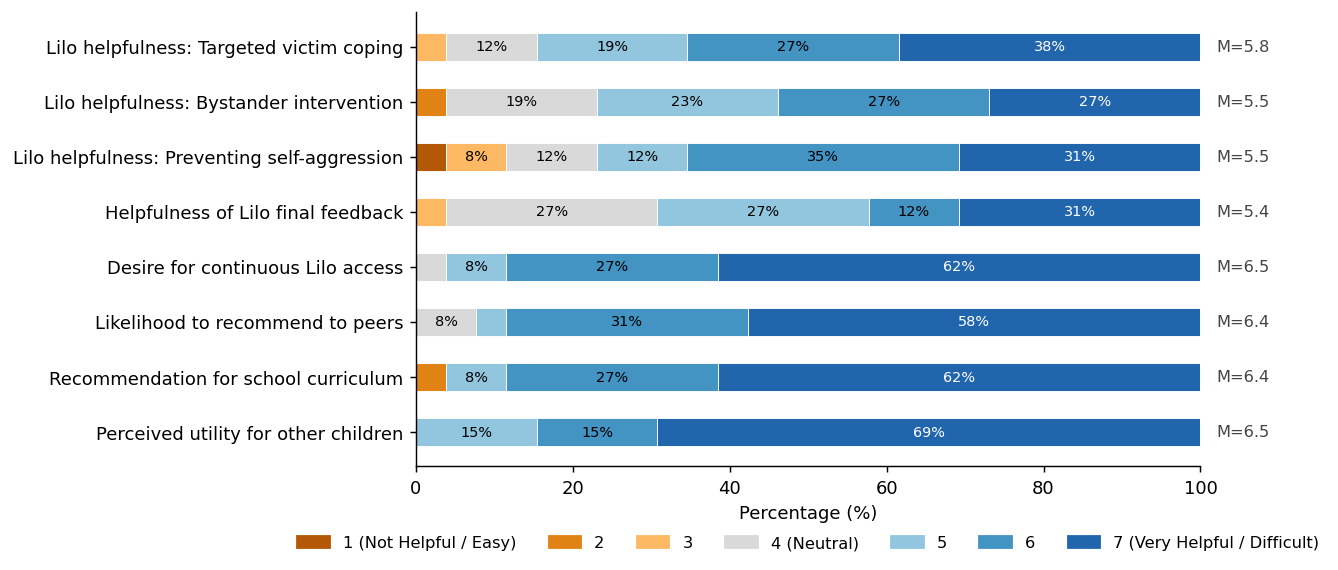}
  \caption{Post-evaluation of Lilo}
  \label{fig:post_lilo}
\end{figure*}

The last part of the post-survey included questions to assess participants' perceived future value of their experience with Lilo. 
Overall, as presented in Figure~\ref{fig:post_lilo}, participants evaluated the Lilo experience positively, but the strength of agreement varied by topic. Willingness to continue using and recommending Lilo to peers drew strong, homogeneous agreement, while its perceived usefulness for specific future coping situations was rated positively but somewhat lower and more variable.
Specifically, concerning Lilo's perceived usefulness, participants expected the Lilo experience to help them cope with future situations in which unpleasant behavior was directed toward them (M=5.85, SD=1.19), and to help them respond when another child was treated unpleasantly (M=5.50, SD=1.30). Participants also expected the experience to help them avoid behaving unpleasantly toward other children (M=5.54, SD=1.56).

Referring to willingness to use Lilo, participants expressed a strong desire to have the option to use it (M=6.46, SD=0.81), and reported that they would recommend it to their friends (M=6.38, SD=0.90). They similarly reported that they would recommend using Lilo in school life-skills lessons (M=6.38, SD=1.10). Participants expected access to Lilo to be helpful for other children (M=6.54, SD=0.76). This item received the highest mean evaluation, and all participants rated it between 5 and 7.
Lastly, one item was designed to learn about participants' perceptions of Lilo's end-of-conversation feedback. This item, while overall positive, received the lowest mean score (M=5.38, SD=1.30).

\subsection{Interview Findings}
After the intervention with Lilo and the post-survey, we conducted interviews to learn about participants' experience with Lilo, including their perceptions of the characters, main functions, and how useful and enjoyable they found it. 
In the following sections, we present a detailed analysis of 17 interviews. We organize the findings around two broad topics: (1) appraisals of the system - how participants judged Lilo, the characters, and the guidance they received, and (2) perceived value for themselves and others - what participants believed the experience gave them, and where and by whom it should be used.

\subsubsection{Appraisals of the system}
\paragraph{Perceptions of Lilo}
Most participants had a positive perception of Lilo (the mentor) and gave several reasons for it. We also note that most participants referred to Lilo in the feminine, and we follow their usage in this section. The reasons they gave were of two kinds. The first was her manner. Participants described her as cute, friendly and kind, and several characterized her as a girl who cares about everyone and does not want anyone to fight. 
The second was what she did for them during the conversation, and the way she went about it. Participants said she gave them ideas, told them what to pay attention to, and directed them to the other side's feelings. 
Several described her guidance as responsive rather than general, responding to almost every message and telling them which moves were helpful and which were not. 
The same detailed, instructive style was described positively and, by some, as annoying. The positive perception is illustrated by P1 (g):
\begin{quote}``\textit{She sounds very cute, and that she tries very hard to help all the time, and
that she's a good friend. That she's always trying to give instructions on how to do good things,
and not to fight, and to be good friends.}''
\end{quote}
\noindent In contrast, P2 (b) described the same attentiveness as intrusive:
\begin{quote}``\textit{I wrote something and then she told me to pay attention to all
kinds of other things, and that's kind of annoying.}''
\end{quote}
As for Lilo's perceived age, while a few participants placed Lilo among themselves, describing her as a child or as an excelling pupil, 
most perceived her as an adult. 
The figures they reached for were figures of authority and care. Participants likened her to a parent, to a grown-up in the group who watches over it without taking part, and to a teacher giving advice. 
A smaller group described Lilo as not sounding like a real person at all. 
These participants called her a robot or said the exchange felt computer-like. 

\paragraph{Traffic-Light model}
\label{subsec:trafficlight-interview}
Participants found the traffic-light concept useful, but in most interviews they did not spontaneously recall that it had been introduced and defined at the start of the session. Once the interviewer mentioned the colors, they engaged with the model readily and described it in their own words. Their descriptions represented a process in which they were making (or not) a progress, with a positive (green) or negative (red) direction, and yellow somewhere in between, as explained by P12 (g): 
\begin{quote}``\textit{.. it really helped me understand that green also means I'm in a good direction, red means I'm not in a good direction, and yellow means I'm in the middle.}''
\end{quote}

\noindent
P15 (g) described in her words how she perceived the traffic-light idea:
\begin{quote}``\textit{... if you overdo it and it starts being hurtful, then they kind of tell you stop. And you can understand that you need to shift down a gear, to look at what you wrote, to try to find maybe there's a way, that maybe the concept of what you wrote is right, but maybe in other words..}''
\end{quote}


Participants then described how each level worked for them. Red was understood as a signal to stop, reserved for messages that were in the wrong direction or had become offensive. 
Participants described how red messages called them to stop and think, and how they used it to pause and re-evaluate the situation, as described by P15 (g): 
\begin{quote}``\textit{When she gave me red lights, I understood I need to pause for a second and think, that maybe what I said needs changing, maybe I need to add something.}''
\end{quote}

Yellow occupied the middle of the scale, neither approving nor rejecting a message. Participants
often associated this middle ground with being asked to reconsider the other side, although they
attributed the prompt to Lilo rather than to the color itself. They described being moved off a
single position in the dispute and toward the character they had not been supporting. 
P7 (g) described it more generally, not specifically in the context of the experienced situation:
\begin{quote}``\textit{Even when I think I'm right, sometimes the other side doesn't agree, and you have to respect that.}''
\end{quote}

Lastly, green functioned as confirmation to continue. Participants described it as a sign that they were on
the right track and that their message had been well judged, and several experienced it as
encouragement, recalling that Lilo told them they had done well. 

\paragraph{Scenario and characters authenticity}
Participants were asked about two aspects of the experience's authenticity, the children's characters and the scenario they discussed, a conflict about a birthday party. First, most participants placed the characters within a year of their own age. 
A few judged them slightly older, and a few younger. 
Beyond their age, most participants said the characters sounded like real children. Participants pointed to several aspects of what made them sound real, with arguing and being caught in a conflict as one such example, which they described as something children do.
P14 (b), for example, said the characters reminded him of children who do not get along, and confirmed that this was precisely what made them feel real. 
A few participants located the realism in how the characters talked, for example, referring to the style in which they wrote to each other. One participant repeated a line back as the kind of thing children say, quoting Sagi asking why everyone was making a drama when he only wanted to invite his own friends. 
For example, P12 (g) described how Sagi and Reut sounded to her:
\begin{quote}``\textit{They really sounded to me like real kids who are in the middle of a fight and really need to solve it.}''
\end{quote}
On the other hand, several participants explained why Sagi and Reut did not sound like real children. First, they referred to the way the characters spoke, saying they did not talk like people and that the artificial intelligence was apparent. 
P9 (g), for example, put it as \textit{``they don't talk like people, you can tell it's AI, their speech isn't like how I talk with my friends''}.
A related observation concerned what the characters failed to do. One participant reported using the word ``empathetic'' in the conversation, and noted that the characters let it pass, where children her own age would have remarked on it. 
\begin{quote}``\textit{
P15: For example, when I wrote ``empathetic'', other kids wouldn't have reacted the way they did,
they'd have reacted a bit differently [...] because usually in conversations people don't say the word
``empathetic''.
\\
Interviewer: What do you mean? They'd have pointed out that you said ``empathetic''?
\\
P15: Yes, they [real children] would have said to me that it sounds grown-up, something like that.}'' 
\end{quote}
Second, participants referred to inflexibility. One participant said the characters were too stubborn, explaining that with real friends he can usually bring people around, and here he could not. 
One participant said they did not sound especially real to her, noting that she knew throughout that it was a simulation, which may be what kept her from treating them as real. 

Finally, a few participants questioned the premise rather than the characters themselves. One participant, who said the characters strongly reminded him of real children, added that girls in his class do not behave as Reut did, explaining that boys and girls in his class are not really friends. 
Another found it odd that Sagi would state a boys-only preference in the way he did, and that children would have phrased it differently. 
This concerns the scenario as much as the characters, and participants also addressed the scenario directly.
Most considered it realistic, reporting that they could imagine a similar conflict arising in their own group chats, and in several cases that they had experienced something similar themselves. 
A few participants did not see the conflict as reflecting their current social world, whether because it does not occur among the children they know or because they associated it with younger children.

\subsubsection{Perceived value and future directions}
\paragraph{Learning process}
During the discussion of their general feelings about the experience with Lilo, many participants spontaneously described Lilo as a place to practice and framed the session as preparation for situations they might encounter in their own group chats. 
Others valued being able to try something out and see how it was received, as exemplified by P13 (g):
\begin{quote}``\textit{It's really fun, and it also feels like I can practice a bit. If I want to say something and I'm not sure whether it might hurt, I feel like I can say it there [...] and then test it, see.}''
\end{quote}
Participants also explained what made practicing possible. Several referred to the absence of social cost, contrasting the simulated group with their own, where a badly judged message can be held against them long afterwards, and noting that the characters would neither judge them nor carry the exchange forward (beyond the session), as explained by P15 (g):
\begin{quote}``\textit{On WhatsApp with people you know, if you write something bad, people can hold it against you, even a week later. But with an AI, people won't judge you, no matter what you write.}''
\end{quote}

Several said that if a similar situation arose in their own group chats, they would now know how to behave, and that the session had shown them how such situations can be handled. 
Along the same lines, participants conveyed a sense of increased self-efficacy, saying that they had learned to make peace between friends, 
and that the conversation would not have ended well without them. 
P1 (g) described her feelings after using Lilo:
\begin{quote}``\textit{Today my friends had a fight, and I couldn't make peace between them. But now that I've helped someone on WhatsApp, at least someone who isn't real, I think I've learned better
how.}''
\end{quote}

A few participants went beyond the use Lilo was designed for, describing it not as a place to practice but as a resource to consult about real situations as they arose, whether before replying in a conversation or while an argument was unfolding. 
For example, P8 (g) said that if she had argued with a friend and was still angry, she would want to talk it through with Lilo and work that case out.


\paragraph{Perceived impact}
Most participants described the session as an enjoyable experience, with a few attributing their enjoyment to having helped others. 
Relatedly, participants wanted access to Lilo beyond the session, consistent with the survey results reported in Section~\ref{subsec:postsurvey}, and what they imagined doing with it differed by setting. At school they described it as something a class would do together, for example, as part of a recurring activity with a new scenario each time. 
At home, by contrast, participants described personal and on-demand use, wanting Lilo on their own
phone, 
treating it as something to turn to when a real conflict arose, 
or using it daily so that it would keep teaching them things. 

Participants were asked whether Lilo might help children who behave unpleasantly in group chats, those
on the receiving end, or both. Several answered both, while others emphasized one side. 
Their explanations of the children who behave unpleasantly were consistent. Participants described the
problem as a failure to notice rather than an intention to hurt, saying that children do not always
realize what they are writing or that it is insulting, and that a phone makes this easier because the
sender cannot see the other person's reaction. 
Two participants went further and described the remedy as experiencing the other position rather than
being told about it. 
For children on the receiving end, participants expected Lilo to help them know how to respond without
escalating the exchange, or how to 
stand up for themselves rather than stay silent. 
P17 (b) referred to both roles: 
\begin{quote}``\textit{... whoever is treated unpleasantly, then he'd be able to know how to act. And whoever behaves unpleasantly can understand that what he's doing isn't good, and if it were done to him, then he would feel it.}''
\end{quote}

Finally, participants also described what such use might change, at three scales. At the level of their own
conduct, most said they could apply the traffic-light model beyond the session, most often as a rule
about pausing before sending a message, or to take a disagreement into a private conversation rather than airing it in the group. 
At the level of the group, participants expected fewer conflicts if other children came to recognize the effect of what they write. 
One participant extended this to the platform itself, as explained by P15 (g):
\begin{quote}``\textit{I really think it could change conversations on WhatsApp, if WhatsApp did what Lilo does and put in a traffic light. Stop. Stop for a second, look at what you wrote, try to rephrase it.}''
\end{quote}

\paragraph{Suggested improvements}
Toward the end of the interview, participants were asked what they would change in Lilo. Most of their suggestions concerned the scenario. Participants proposed other topics for conflict scenarios, some more difficult than the one they encountered, including social exclusion and shaming, and some of a similar kind, such as an outing to a movie. 
Their motivations for suggesting other scenarios varied. Some wanted variety because a single fixed story would not sustain repeated use, which several of them said they would want. 
Others wanted a more challenging conflict because it would leave them more room to intervene, as explained by P16 (g): 
\begin{quote}``\textit{The topic was meh. If it had been a topic about everyone ganging up on someone, then it would have been more [possible] to intervene. But, it's a birthday, so at the end of the day it's your day, you decide what to do.}''
\end{quote}

A second group of suggestions concerned Lilo's guidance itself. Two participants asked for more concrete help, either a button that would let them ask Lilo what the problem was and what to write, or ready phrasings they could send. 
Another topic concerned the conversation's composition. 
Participants wanted more characters, both because a real class group is far larger, with one participant contrasting the three-person chat with the twenty-four children in her own group, and because, with only Sagi’s and Reut’s opinions positioned far apart, there was nothing in between.
Finally, several participants referred to interface-related issues, such as not being able to send consecutive messages before a character replied, as in a real group chat. 

\section{Discussion}\label{sec:discussion}


\subsection{Practicing on a conflict that is not real}

Children reported significantly greater perceived ability to cope with unpleasant behavior on WhatsApp after their experience with Lilo, both when the behavior was directed at them and when another child was targeted. We found no comparable change in how easily they felt they could identify such behavior, which is unsurprising in retrospect, since the intervention was not designed to affect it.

We use self-efficacy theory to further discuss this finding. Bandura describes four sources of efficacy beliefs - performance accomplishments, vicarious experience, verbal persuasion, and physiological states \cite{bandura1977self}. Two of these are present in Lilo, and both surfaced in what participants told us. Children wrote their own messages into the conflict, which is a performance accomplishment, and several described themselves as the ones who had ended the disagreement, saying they had helped, made peace, or found a solution that everyone would accept. Verbal persuasion was also present, as children described how Lilo told them when they had done well. Our findings are in line with prior work that placed users inside the exchange, where self-efficacy rose \cite{lin2024imbue} and learned strategies were used more
\cite{shaikh2024rehearsal}. Our findings suggest this direction holds promise for children as well, in the digital conflicts they face themselves.

Beyond self-efficacy, participants also valued not only success but also the safety of failure. Several described Lilo as a place to practice and contrasted the simulated group with their own. In a real class group chat, a poorly judged message can be held against them even a week later, while with Lilo, they said, nobody would judge them whatever they wrote. 
Earlier work asked children, teens, and educators what they would want from a chatbot that supports them through cyberbullying and online conflict, and being non-judgmental was one of their answers \cite{piccolo2021chatbots, zou2024social}.  
Our participants suggest this property does real work. They not only found Lilo comfortable, but also said it let them try things they would not try in their own group.

\subsection{What makes a simulated peer credible to children}

To increase children's engagement with the chat simulation, we aimed to create a group chat that would resemble their usual experience of digital communication with friends. Much of that resemblance rests on the peer characters, and on whether children read them as real. From what the participants told us in the interviews, we identified three factors on which they judged the characters' credibility. Two of the three worked in both directions, and the same feature that convinced one child raised doubt for another.

The first factor is the characters' \textit{register}, the way they wrote to each other. Some participants pointed to it as evidence that the characters were real, describing the style in which Sagi and Reut wrote to each other, such as \textit{``making a drama.''} Other participants pointed to the same feature to explain why it was clear that the characters were AI-based, saying they did not talk like real people and that their vocabulary was too mature for children their age. This observation supports prior work, which showed that chatbots' register characteristics influenced their perceived credibility \cite{chaves2022chatbots}. With children specifically, prior work found that AI-generated speech was associated with perceptions of the AI agent's experience, i.e., whether it seems capable of feeling emotions or sensations \cite{xu2025makes}.

The second factor is the characters' \textit{persona}. Participants described several elements of the characters that together point to a persona. They described Sagi as insulting and Reut as offended, and that they sounded like children who do not get along. One participant (P7, g) went further and assigned them motives, describing Reut as a girl who wants attention, and Sagi as someone who treats everything as a joke. Such descriptions concern who the characters were, and not only how they expressed themselves. 
These accounts are in line with how Lilo was built. Each character was given a stable identity, a communication style, and behavioral tendencies in advance \cite{tseng2024two}, and what the participants told us suggests that this layer reached them, since they reconstructed the characters in their own words.
Participants also pointed to a single trait, in our case the characters' stubbornness. For several of them, it was evidence that the characters were real, while another said they were too stubborn, explaining that with real friends he can usually bring people around and here he could not. In the context of persona, these accounts show the extent to which such traits depend on calibration.

The third factor is the characters' \textit{conduct in the conversation}. Here, we refer to whether the characters' responses, not just their words, felt appropriately human. Participants referred to this in two forms: how the characters conducted themselves toward one another (for example, Sagi's confrontational tone toward Reut), and how they reacted to the participant. Several participants pointed to specific moments where a character failed to react as a real peer would, such as not remarking on the participant's use of a grown-up word like ``empathetic.'' Focusing specifically on this AI-participant dimension of conduct, prior work supports the observed impact of AI-agent conversational conduct on the way it is perceived by users. Exploring support-seeking conversations with a mental health chatbot, Rheu et al. found that when the chatbot's responses violated users' expectations of an appropriate reaction, users trusted the chatbot less and rated the overall interaction more negatively, even when the actual content of its response was accurate \cite{rheu2024chatbot}.


To summarize, credibility is not a binary property, and it is not one to maximize. In two of the three factors, ``failure'' was possible in both directions. Stubbornness was praised as the way children behave and criticized when there was too much of it. The characters' conduct was faulted both for not reacting and for reacting too much. Beyond that, maximum human-likeness was not the goal. Children found the characters credible while knowing they were simulated, and held both views without feeling that they contradicted each other.
In a text-based environment, where the non-verbal cues that guide face-to-face communication are missing 
\cite{walther1992interpersonal, daft1986organizational}, building a simulated peer for children is a matter of calibration and context, and not only of linguistic accuracy. This is in line with prior work showing that children distinguish between people and machines and value that distinction \cite{hiniker2021can}, and with what children and educators themselves asked for, a system that is human in character without pretending to be a person \cite{piccolo2021chatbots, zou2024social}.

\subsection{A place to practice or a source of help}

Participants had several suggestions and requests when asked what they would change or add to Lilo. Some of these requests reflect a general desire for \textit{on-demand help}. 
Beyond having a place to practice pre-defined scenarios or receiving situationally grounded ideas for how to proceed in a conflict, participants imagined Lilo as a character to consult on ongoing real conflicts,  one that would also supply ready-made responses they could use directly. However, Lilo is deliberately not designed this way. It offers ideas, suggests other points of view, and asks questions, ultimately functioning as a coach rather than a counselor. 

A space to practice and an on-demand counselor are two different systems that serve different goals, and both have a place. Participants' wish for the second is understandable, and it points to the tension between them. In the context of autistic adults, prior work has seen the two as complementary, and proposes that future work explore how to combine simulation with real-time assistance within a single system \cite{haroon2024twips}.
However, considering the population we studied, neurotypical children, and the context, teaching positive digital communication, we suggest keeping the two separate. A similar concern was raised in another domain, where students practicing high school mathematics with an assistant that gave finished solutions later scored lower than students who had never used one, while an assistant that gave hints instead did not produce this effect \cite{bastani2025generative}.

Such a separation is derived from the participants' responses as well. The value they described came from the fact that failing to communicate appropriately carried no social cost, and that no one would remember the next day, or for weeks, what they had written. An on-demand assistant that works inside a real conflict brings that cost back, and adds a risk of its own. In a real situation, an inaccurate or poorly judged suggestion from an LLM-based assistant reaches a child at the point where such an error can cause harm. This is unlike a controlled, pre-defined scenario, where learning, including making mistakes, is an essential part of the process. We do not claim, therefore, that an on-demand assistant is redundant, but that it seems preferable to keep the two distinct.

\section{Limitations and Future Work}\label{sec:limitations}
While our study advances understanding of how children experience practicing digital communication with an LLM-based system, several limitations apply.
Our study consisted of a single session, in which we measured children's experience and their view of whether Lilo would help them in the future, not how the intervention affects their communication in practice. A longitudinal design would be better suited to that question. A single session nevertheless supports an exploratory first study, allowing us to elicit information (such as children's perceptions of the system and what might improve it) that contributes to the HCI literature and informs a future version of Lilo, which could then be used in longitudinal research.

Several additional limitations relate to the study design. While we observed a significant effect in children's self-efficacy, it may be specific to the scenario the children practiced. In a randomized trial of simulation-based communication training with adults, gains in skill transferred to new situations, while gains in self-efficacy did not \cite{lin2024imbue}. Furthermore, our scenario used a fixed gender pairing: Sagi, the boy, always refused to invite Reut, the girl. A better design would counterbalance these character roles across sessions, and would also counterbalance the gender of the participant.

In addition, we did not have a control group to explore alternative explanations for the difference observed in participants' self-efficacy measures. These measures were also administered aloud by the experimenter rather than privately. We cannot rule out that the change reflects taking part in a study, an hour spent thinking about conflict on WhatsApp, or a wish to report positively, rather than the intervention itself. A controlled design is the appropriate next step, and comparable systems have been evaluated that way \cite{shaikh2024rehearsal, lin2024imbue}.

Lastly, our sample was homogeneous; for example, all our participants owned smartphones, and most of them had highly educated parents. Such homogeneity limits how far our findings generalize, though our aim is transferable design insight rather than broad generalization. Children with diverse experience, either with phone use or group-chat use, as well as those whose families have different backgrounds, may relate to Lilo differently.

Considering these limitations, several directions follow. The questions we could not answer quantitatively deserve dedicated measures. We did not measure the extent to which the characters were perceived as realistic, nor how that perception related to children's engagement in the conversation, and both could be operationalized and tested directly. A longitudinal study would address the limitation of the single-session design. Deploying Lilo in a classroom across repeated sessions and varied scenarios would make it possible to ask whether children's communication in their own group chats changes over time, rather than whether they believe it might. Finally, Lilo is designed for Hebrew, and its language is calibrated to the digital register of Israeli children. Extending it to other languages and societies would test whether the design holds where the norms of peer communication differ, and would widen access to it.

\section*{Ethics and Privacy Statement}
Our work points to the potential of AI-based conversational systems to help children practice challenging digital-communication situations so that they are better prepared to respond in real life. This potential carries risks common to AI systems used with children, including over-reliance on the system and erosion of independently developed communication skills, as well as the risk that a hallucinated or poorly judged model response could teach a child an inappropriate way of handling a real situation. A related risk is that such practice may take place without a parent's or teacher's involvement, whereas embedding it in settings with adult guidance, such as a classroom activity, could make the learning experience more holistic. We built in several mitigations: a real-time safety layer that intercepts harmful language, and a design deliberately restricted to a bounded practice scenario rather than an on-demand advisor for real, ongoing conflicts, a use case we argue requires separate safeguards beyond our system's scope. Finally, because our system relies on a commercial, cloud-hosted LLM, it raises the same data-privacy considerations as other child-facing AI tools; we mitigate this in part through a pre-defined, scripted scenario that invites children to role-play rather than share personal information, and future versions could further reduce this risk with a locally hosted model.

\bibliographystyle{ACM-Reference-Format}
\bibliography{ref}

\appendix
\section{Study procedure instruments}
\label{sec:instruments}
The following instruments can be found in the link below:
Introductory text, pre- and post-survey, exit-interview protocol:

\url{https://osf.io/x4hqe/overview?view_only=4ed01a88718b4b7eae0a3eddd7e1fcfd}

\section{Coping strategies for insult situations}
\label{sec:coping}

\begin{figure*}[]
  \centering
  \includegraphics[width=\textwidth]{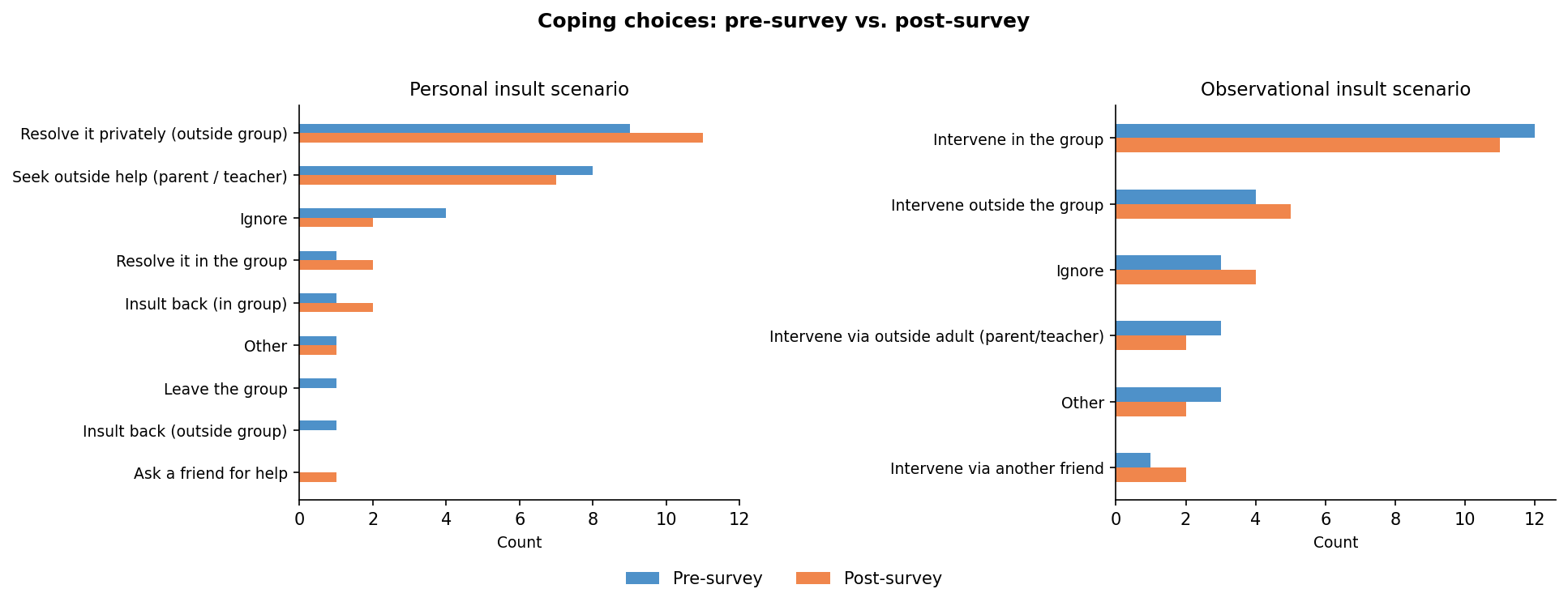}
  \caption{Coping strategies for insult situations, either personally (left), or observing someone else being insulted (right)}
  \label{fig:coping_strategies}
\end{figure*}

\section{Codebook Used for Thematic Analysis}
\label{sec:codebook}

The exit interviews were analyzed using a three-level coding hierarchy. For concision, the codebook
below reports the first two levels and provides a brief description of the content captured by each
code. Descriptions of second-level codes reflect the corresponding third-level codes used during
analysis, where applicable. Third-level codes are not shown.

\begin{table*}[t]
\centering
\small
\caption{Codebook used for thematic analysis. The codebook was applied to the exit interviews included
in the analysis; the two pilot sessions were used to develop the codebook and are excluded.}
\label{tab:codebook}
\begin{tabularx}{\textwidth}{@{}l l X@{}}
\toprule
\textbf{Main category} & \textbf{Subcategory} & \textbf{Description} \\
\midrule

Educational approach & Perception of Lilo &
Perceptions of Lilo, including its perceived age, whether it sounded like a real person, positive or
negative impressions of the character, and its friendly, non-instructional learning tone. \\
\addlinespace

Educational approach & Traffic-light model &
Perceptions of the traffic-light model and its educational mechanisms, including reframing, real-time
feedback, self-reflection, and positive reinforcement. \\
\addlinespace

Authenticity & Characters authenticity &
Perceptions of whether the peer characters sounded like real children or did not, and whether their
portrayal was age-appropriate. \\
\addlinespace

Authenticity & Perception of characters &
Participants' impressions of the peer characters as individuals, including their perceived
dispositions and behavior. \\
\addlinespace

Authenticity & Scenario authenticity &
Perceptions of whether the scenario reflected real-life situations, including personal identification
with the topic and whether the scenario did or did not resemble real-life experiences. \\
\addlinespace

Authenticity & Authentic self-expression &
Participants' perceptions of their ability to express themselves authentically during the
interaction. \\
\addlinespace

Learning process & Safe practice &
Perceptions of Lilo as a practice environment, including emotional safety and interaction with AI as a
non-judgmental space. \\
\addlinespace

Learning process & Self-efficacy &
Participants' perceptions of their self-efficacy in the context of the experience. \\
\addlinespace

Learning process & Relevance to personal experience &
Connections between the experience and situations encountered in participants' real lives, including
negative ones. \\
\addlinespace

Learning process & Motivation and engagement &
Participants' engagement with the experience, including experiences of loneliness or responsibility
burden during the interaction. \\
\addlinespace

Attitudes & Perceived impact &
Perceived impact of the experience, including conflict management, application of the traffic-light
model, changes in attitude, potential for future use, and changes in communication discourse. \\
\addlinespace

Attitudes & Positive satisfaction &
Positive expressions of satisfaction with the experience. \\
\addlinespace

Suggestions for improvement & --- &
Suggestions for changes or improvements to Lilo or the experience. \\

\bottomrule
\end{tabularx}
\end{table*}

\end{document}